# Enabling the Reuse of Personal Data in Research: A Classification Model for Legal Compliance

Eduard Mata i Noguera[1*], Ruben Ortiz Uroz[2], Ignasi Labastida i Juan[1]

[*]eduardmatainoguera@ub.edu

Corresponding autor adress: Carrer de Baldiri Reixac, 2, 1a planta, Les Corts, 08028 Barcelona

[1]Unitat de Recerca, CRAI, Universitat de Barcelona, Barcelona, Spain

[2]Serveis Jurídics, Universitat de Barcelona, Barcelona, Spain

**Abstract:** Inspired by a proposal made almost ten years ago, this paper presents a model for classifying personal data for research to inform researchers on how to manage them. The classification is based on the principles of the European General Data Protection Regulation and its implementation under the Spanish Law. The paper also describes in which conditions personal data may be stored and can be accessed ensuring compliance with data protection regulations and safeguarding privacy. The work has been developed collaboratively by the Library and the Data Protection Office. The outcomes of this collaboration are a decision tree for researchers and a list of requirements for research data repositories to store and grant access to personal data securely. This proposal is aligned with the FAIR principles and the commitment for responsible open science practices.

## 1. Introduction

Open Science practices are fostered by institutions and research funders as a way to make research more collaborative, transparent and closer to society. Among these practices we find the effort to make research data useful for reuse. To achieve this goal the FAIR principles were developed (Wilkinson et al., 2016) and consolidated (Jacobsen et al., 2020). The implementation of these principles to research data fosters their opening but also the need to open their metadata when data cannot be shared publicly. When managing personal data from research activities we find this later situation: Data cannot be openly shared.

A decade ago, researchers at Harvard proposed the idea of tagging personal data to provide researchers with a tool to know how to share this kind of data (Sweeney et al., 2015; Bar-Sinai et al., 2016). That project was created following US applicable laws what required an adaptation to be used in other legal frameworks. Year later, DANS, the Dutch national centre of expertise and repository for research data, began adapting the model (von Stein, 2017; Baxter et al., n.d.) within the European General Data Protection Regulation (GDPR) framework (Regulation - 2016/679 - EN - Gdpr - EUR-Lex, n.d.). Although this project was never completed, there are certain projects that came out of this idea (Sansone et al., 2017) (Alter et al., 2020). It was also the predecessor of ours.

The library at the University, known as CRAI (Centre de Recursos per l'Aprenentatge i la Investigació), currently provides support to manage research data, especially in developing data management plans and in publishing data in the consortium repository, CORA.Repositori de Dades de Recerca (CORA.RDR). Until now, this repository doesn't allow the deposit of personal data and researchers often ask how to manage and keep personal data safely. These were the two main reason to develop the current work and continue what DANS started inspired by the American Datatags. Initially we used the GDPR as the legal foundation to build our tools but when we invited the Data Protection Office, we focused on the national implementation of the GDPR because the national law can introduce differences between the Member States of the EU. The European Regulation allows the Member States of the EU to complete its provisions, what was made in Spain through the Organic Law 3/2018 of December 6, on Personal Data Protection and the guarantee of digital rights (Official Gazette of Spain number 294 of 6 December 2018.).

This work can be divided into two key phases. The first phase involved designing a decision tree (see Figure 1) and defining the data tags, providing researchers with a practical tool to assess the nature of the data they handle. This phase also demonstrated that the FAIR principles (Findable, Accessible, Interoperable, and Reusable) can still be upheld even when certain data must remain closed due to security and privacy concerns. The decision tree serves to uphold the principle of "as open as possible, but as closed as necessary", challenging the misconception that non-open data cannot adhere to FAIR principles. We try to show that open science must be done responsibly and when it's necessary to close sensitive data. The second phase focuses on the implementation of the necessary security and precautionary measures in research data repositories. Our next step is to work on integrating these data tags into the CORA.RDR, ensuring that the appropriate safeguards are in place to protect sensitive data while maintaining its accessibility for research purposes.

## 2. Legal Framework

The main legal framework for the protection of personal data, including in research, is the General Data Protection Regulation. Though the GDPR sets a very high standard for data protection, it also contains important provisions that accommodate the unique needs of scientific research and balance the protection of personal data with the advancement of knowledge.

One of the important features of the GDPR, in research, is its flexibility. Article 9 explicitly recognizes the importance of scientific research and allows the processing of special categories of personal data, under certain conditions, without explicit consent. For instance, personal data may be processed when research is in the public interest, provided that appropriate safeguards, such as pseudonymization or anonymization, are implemented to reduce risks for individuals. The GDPR also allows personal data collected for one purpose to be reused for compatible research purposes, provided that such use respects the principles of data minimization and purpose limitation, as outlined in Article 5.

The GDPR specifically addresses Special Categories of Data in Article 9. These include data revealing racial or ethnic origin, political opinions, religious or philosophical beliefs, or trade union membership, as well as the processing of genetic data, biometric data for the purpose of uniquely identifying a natural person, data concerning health, or data concerning a natural person's sex life or sexual orientation. The processing of such data is generally prohibited unless specific conditions are met, such as obtaining explicit consent from the data subject or if the processing is necessary for scientific research purposes based on Union or Member State law, subject to appropriate safeguards to protect the rights and freedoms of the data subjects.

In Spain, the GDPR is complemented by Organic Law 3/2018, of December 5, on Personal Data Protection and the Guarantee of Digital Rights (LOPDGDD), which fills critical gaps and introduces more flexible measures in certain areas. The LOPDGDD tailors the GDPR to the Spanish context, providing detailed regulations for processing health data in scientific research.

The LOPDGDD aligns with the GDPR by allowing the processing of health data for research without explicit consent under specific conditions, such as when the research is carried out in the public interest. However, it imposes additional safeguards, including stricter requirements for pseudonymization, encryption, and access control. Moreover, the LOPDGDD mandates that data protection impact assessments (DPIAs) be conducted for research projects involving sensitive data in the cases laid down by Article 35 of Regulation (EU) 2016/679 or in those established by the supervisory authority.

One area where the LOPDGDD introduces further specificity is in the retention and reuse of data for research purposes. While the GDPR allows data to be reused for compatible purposes, the LOPDGDD explicitly requires that researchers establish clear protocols for ensuring compliance with data minimization and proportionality principles. It also defines additional restrictions for certain types of research data, requiring explicit legal or ethical justifications to override the rights of individuals.

Recent European regulatory developments such as the Data Governance Act (DGA) (Official Journal of the European Union number L 152/1 of 6 June 2022) are highly relevant to the broader debate on data reuse and governance. However, their direct applicability to the framework proposed in this study requires careful examination. Chapter II of the DGA regulates the re-use of certain categories of protected data held by public sector bodies. Nevertheless, Article 3(2)(c) explicitly excludes from its scope "data held by cultural establishments and educational establishments." Given that universities fall within the category of educational establishments, the provisions of Chapter II are not applicable to the context addressed in this work.

Chapter III of the DGA establishes requirements for data intermediation services, which could, at first glance, appear relevant to non-commercial research data repositories. Article 2(11)(d) includes within this concept data sharing services offered by bodies that do not aim to establish commercial relationships. However, Recital 29 of the DGA clarifies that repositories whose purpose is to enable the re-use of scientific research data in accordance with open access principles should not be considered data intermediation services within the meaning of the Regulation. Consequently, the framework presented in this study does not fall under the scope of Chapter III of the DGA.

With regard to the European Health Data Space (EHDS) Regulation, it should be noted that its application is deferred until 26 March 2027, with several provisions becoming applicable at a later stage. Furthermore, key aspects related to the secondary use of health data require further development at the national level by Member States of the European Union. In the absence of implementing national legislation and authoritative guidance from competent bodies, it is not yet possible to adequately assess the concrete impact of the EHDS Regulation on the framework proposed in this study. For this reason, the EHDS has not been incorporated into the present analysis.

## 3. Methods and Procedures

The reason for developing this work has been firstly the need for a standardised procedure that facilitates the reuse of research data to contribute to responsible open science, where the guarantee of privacy rights goes hand in hand with compliance with FAIR principles. Therefore, our thought process was to investigate previous works such as those previously discussed in Harvard and DANS, and to take up their projects more specifically for the Spanish legal framework. The development of the labels went hand in hand with the development of the decision tree (see Figure 1). It was a process of optimisation of both parts (tags and tree), which has resulted in a tree with a total of 7 possible outcomes. This study seeks to answer two key research questions: What criteria are used in the labeling system to classify data based on its sensitivity, and what specific consequences and precautions must be taken according to the assigned tag? To address these issues, a decision tree (see Figure 1) based on GDPR and LOPDGDD and a table (see Table 1) outlining the consequences and precautions associated with each type of tag has been created during the development of the project. The procedure for the user of the tool consists of reviewing the decision tree to analyse how tags are assigned based on issues related to the nature of the data and their legal use, as well as examining a table of consequences to identify recommended actions and precautions for each type of data tag.

In order to develop this decision tree in a coherent manner, we knew that we first had to separate the data into three large blocks, and once that was done, examine whether it was necessary to make further distinctions within those blocks Within each of the three main categories of data, a fine-grained labelling logic was developed to reflect the gradation of legal obligations and technical safeguards required by both the GDPR and the Spanish LOPDGDD:

1. Block of non-personal data: This group is made up of data sets that do not contain any identifiable

or re-identifiable information. As no person may be linked to the content, no additional restrictions will apply to the processing. Consequently, a single label – blue – was defined to represent full openness and unrestricted reuse, which did not require authentication or encryption measures.

2. Block of personal data but not sensitive: For the data qualified as "personal" under article 4 of the GDPR, but not belonging to special categories, multiple gradations were needed to reflect different conditions of reuse. The green label corresponds to cases where informed consent explicitly allows reuse for compatible research purposes; therefore, access to authenticated users can be granted under standard security measures. The yellow label represents sets of personal data that lack explicit consent for their reuse, but for which the data controller may still authorise secondary use under the principle of compatibility (Art. 5(1)(b) and recital 50 GDPR). This distinction was introduced to operationalize the legal difference between re-use based on prior consent and re-use based on compatibility assessment, a key element of compliance when managing research data repositories.

3. Special categories of the personal data block: The data in Article 9 of the GDPR and the LOPDGDD required the most detailed differentiation. Two subgroups were created according to the nature of the data and the conditions of consent. For data related to health or genetics, the orange label is applied when reuse is allowed only within a defined medical or research specialty, in accordance with additional provision 17(2)(a) of the LOPDGDD. The red label, on the other hand, denotes data sets of the same nature, but without consent for their reuse, requiring a dependence on an alternative legal basis such as public interest in scientific research (section 2a of additional provision 17a of the LOPDGDD (c–d), according to Art. 9(2)(j) GDPR). For other special categories of data (e.g. ideology, religion, sexual orientation), the purple label reflects the logic of the orange label, applying when reuse is allowed only for specific research areas under conditional consent according to Recital 33 GDPR.

This hierarchical labelling structure allows the model to capture the legal reasoning behind each level of data sensitivity and translate it into the corresponding technical safeguards —authentication, encryption and access control — later detailed in Table 1. The decision to maintain six different labels ensures a balance between legal accuracy and practical usability, avoiding both the sub- and excessive classification of research datasets.

## 4. Classification data by tags

As we aim to create a useful and efficient tool for research and technical staff involved in data management, we have tried to find the optimal point between not having too many questions in the decision tree but just enough to be able to correctly classify the data. Likewise, we used the same idea for the creation of each tag, trying not to generate too large a number of tag with specific characteristics for each one of them that could become unmanageable or impractical, but to generate a sufficient number of them to be able to correctly separate different sets of data that would otherwise have to be closed in a more restrictive way. Here is our proposal that guarantees this optimisation:

- Blue tag: Non-personal data.
- Green tag: Personal data. The publication of the dataset needs to indicate (a) whether the participants were informed that the data would be made available to other researchers or (b) whether consent was obtained that the data could be re-used for other research projects in a particular research area by indicating this area.
- Yellow tag: Personal data requiring the intervention of the data depositor (We understand data depositor as the person responsible for the processing of the data). The intervention of the data depositor is required to assess whether the re-use complies with Article 5.1b of the GDPR and Recital 50 of the GDPR.

- Orange tag: Personal data relating to health or genetics where consent for re-use is available under certain conditions. Intervention by the data depositor is required to assess whether the reuse complies with section 2a of additional provision 17a of the LOPDGDD, considering the consent given by the subject for the data to be reused for other research projects in a general area linked to a medical or research speciality.

- Purple tag: Special categories of personal data other than those related to health or genetics, where consent for re-use is available under certain conditions. Intervention of the data depositor is required to assess whether the re-use of the data complies with Recital 33 of the GDPR and Article 9.2a of the GDPR, considering the consent given by the subject that the data may be re-used for other research projects in a particular area of research.

- Red tag: Personal data relating to health or genetics where consent for re-use is not available. Intervention by the data depositor is required to assess whether the re-use complies with section 2c or 2d of the 17a additional provision of the LOPDGDD.

- No tag possible: This is an end of the decision tree that indicates that the nature of the data is so complex that a prior review of the specific case by the Data Protection Officer of each institution is necessary.

The difference between the orange and the purple tag lies in the scope of the consent for re-use given by the participants in the original project. The orange tag refers to medical or research specialities, the purple to other research areas.

The reason for differentiating between these two tags was to avoid a message being displayed at the end of the decision tree explaining the two criteria depending on the type of data being deposited.

## 5. Implementation for research data repositories

One of the goals of our work was to implement the model in actual repositories that could provide open metadata while securing access and storage for research personal data according to Article 32 of the GDPR. To ensure that research data repositories comply with data protection regulations and adequately safeguard research data, we have classified the requirements into four key areas. These areas help determine the necessary safeguards and actions based on the sensitivity of the data:

- Identification and Authentication: Refers to the process of validating the identity of users accessing the data repository. Depending on the sensitivity of the dataset, authentication may not be required (public access) or more complex systems may be implemented, such as repository registration, passwords, two-factor authentication, and even validation by IP address to ensure that only authorised users have access.
- Read and Download Permissions: Establishes who has the right to view or download data from the repository. This ranges from unrestricted public access to permissions granted exclusively to registered users, which in some cases need explicit approval from the data depositor. For more sensitive data, downloading may be encrypted with passwords, or even disabled completely.
- Storage and Transmission: This refers to measures to protect data during storage in the repository and during transmission between systems. This ranges from unencrypted data (for low-risk tags) to the use of advanced encryption algorithms and double encryption for sensitive data. Transmission should always be through secure channels, such as encrypted connections, to prevent unauthorised access.
- Encryption Key Storage: Describes strategies for protecting the keys used to encrypt data. For more sensitive data, the keys must be stored separately from the data in the repository. In highly sensitive cases, a distributed model is implemented, where one key is managed by the repository and another by a trusted third party, ensuring maximum security even in the event of a breach.

| Tag type | Identification and authentication | Read and download permissions |
|---|---|---|
| Blue | Not necessary | Public access without authentication |
| Green | Registration to the repository is required<br>Implementation of access controls (username and password, certificate, second factor authentication)<br>Assigned roles with privilege differentiation | Access by registered users<br>In case of downloading documentation, encrypted with a password |
| Yellow | Registration to the repository and approval by the data depositor is required.<br>Implementation of access controls (username and password, certificate, second factor authentication)<br>Assigned roles with privilege differentiation | Registered users can access the data after authorisation of the depositor.<br>In case of downloading documentation, encryption with a password |
| Orange | Registration to the repository and approval by the data depositor is required.<br>Implementation of access controls (username and password, certificate, second factor authentication)<br>Assigned roles with privilege differentiation<br>Validation according to source IP | Registered users can access the data after authorisation of the depositor.<br>In case of downloading documentation, encryption with a password |
| Purple | Registration to the repository and approval by the data depositor is required.<br>Implementation of access controls (username and password, certificate, second factor authentication)<br>Assigned roles with privilege differentiation<br>Validation according to source IP | Registered users can access the data after authorisation of the depositor.<br>In case of downloading documentation, encryption with a password |
| Red | Registration to the repository and approval by the data depositor is required.<br>Implementation of access controls (username and password, certificate, second factor authentication)<br>Assigned roles with privilege differentiation<br>Validation according to source IP | Access to protected data without permission to download |

| Tag type | Storage and transmission | Key storage |
|---|---|---|
| Blue | Unencrypted | N.A. |
| Green | Storage: simple encryption<br>Transmission: simple encryption<br>Use secure encryption algorithms | Encryption key stored separately from repository data |
| Yellow | Storage: simple encryption<br>Transmission: simple encryption<br>Use secure encryption algorithms | Encryption key stored separately from repository and depositor data |
| Orange | Storage: double encryption<br>Use secure encryption algorithms | One key is stored separately from the data by the repository, and the other key is stored by a trusted third party. |
| Purple | Storage: double encryption<br>Use secure encryption algorithms | One key is stored separately from the data by the repository, and the other key is stored by a trusted third party. |
| Red | Storage: double encryption<br>Use secure encryption algorithms | One key is stored separately from the data by the repository, and the other key is stored by a trusted third party. |

**Table 1.** The blue to red model for tags categorizes datasets based on their risk levels. Datasets with no associated risks fall under the blue tag, while increasing risk levels demand stricter data protection measures and more complex safeguards, with the red tag assigned to datasets of the highest sensitivity and risk.

While it seems that all measures are the same for the orange and the purple label, the difference is in the organizational measure regarding approval as the depositor will have to consider different criteria.

As part of the implementation effort, we have developed an interactive tool (see Figure 2) that puts the decision tree described in section 4 into practice and translates it into a classification interface useful for researchers and data managers. The tool is a web-based application that guides users through each question in the decision model, dynamically displaying contextual explanations, legal definitions (e.g., “personal data,” “special categories of data,” or the requirements of Article 6 of the GDPR), and references to the relevant provisions of the GDPR and the LOPDGDD. The system processes the user's responses and automatically assigns one of seven possible labels, displaying the corresponding technical and organizational measures required for that classification. By structuring the decision process into a sequence of conditional nodes, the framework is understandable to non-experts. This practical application demonstrates that the model can be integrated into a repository workflow: the tool can be incorporated into the deposit interface, so that depositors must complete the decision path before uploading a dataset, ensuring that a label is assigned consistently and that the repository records the legal basis and consent conditions necessary for reuse. Although integration into CORA.RDR is still ongoing, the development of this tool offers a concrete demonstration of the framework's viability and clarifies how it can be applied in real data management work.

## 6. Discussion

The classification of research data using data labels offers a practical and compliant solution for managing sensitive data (i.e. special categories of data according to the GDPR). Each tag provides a specific framework to help researchers and data controllers comply with legal and ethical obligations. The implementation of data labels is essential to properly manage the risks associated with the processing of research data. It also provides a standardized methodology that could facilitate future audits and compliance reviews.

The original Datatags project was created with the idea of being implemented in a Dataverse environment. Our consortium repository uses such an environment, and the project has already been presented for its deployment there. We hope in a short period of time it will be available for researchers along with the decision tree. Our aim is to improve the reuse of research data while keeping personal data safe when needed, following the lemma of as open as possible as closed as necessary.

FAIR data and responsible open science are fully compatible with robust security measures, ensuring the protection of sensitive data while enabling data sharing and reuse. The goal of this work is to provide a standardized tool to facilitate the identification, classification, and subsequent management of research data. Future work includes integrating this tagging system into CORA.RDR, the institutional research data repository.

## 7. Decision tree

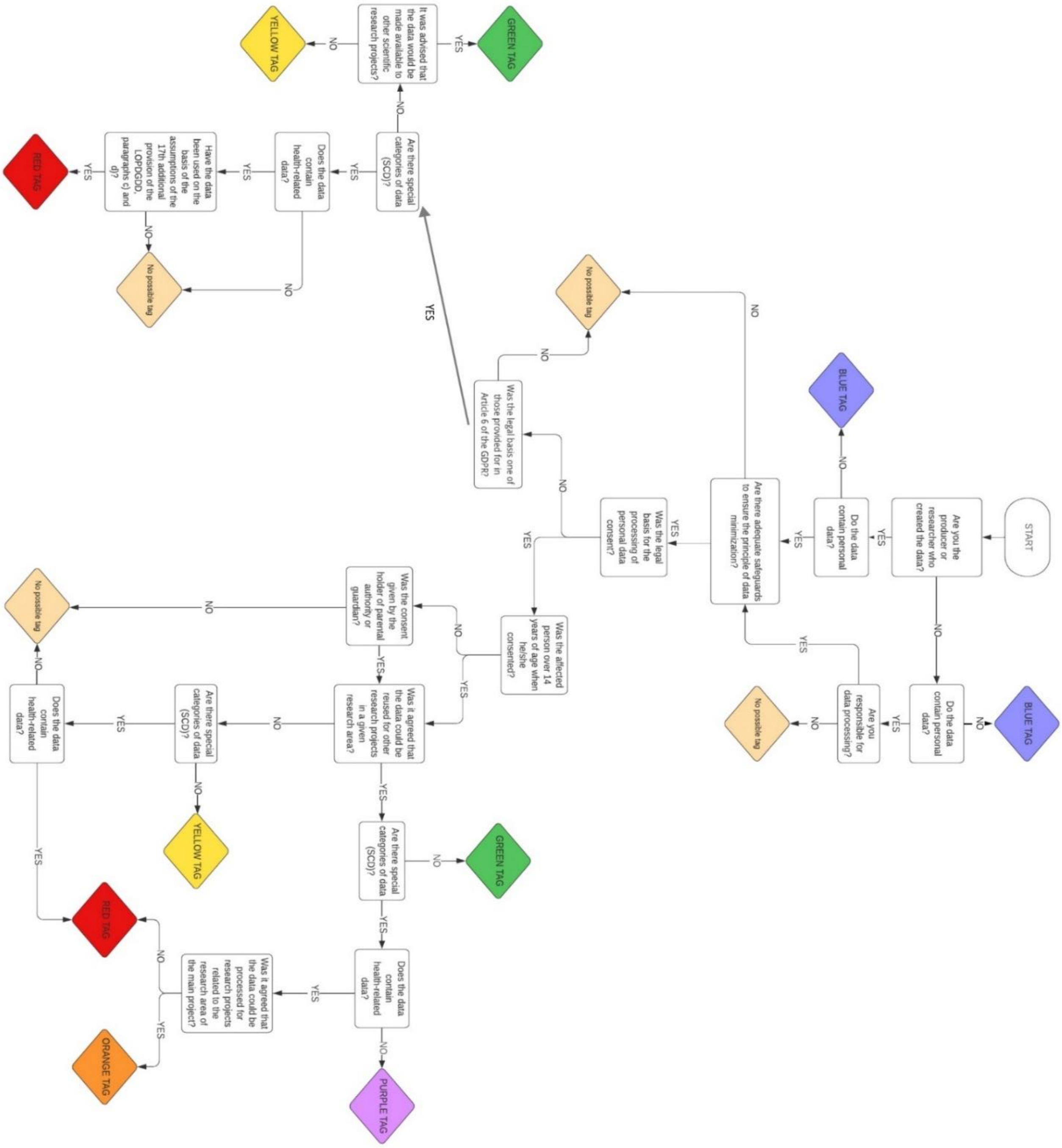

**Figure 1**. Decision tree for the classification of personal data. This diagram guides researchers and depositors in assigning tags to datasets containing personal data based on their conditions for reuse and compliance with the General Data Protection Regulation (GDPR) and Spanish law. The color-coded tags (blue, green, yellow, orange, purple and red) indicate different legal bases and limitations for the secure storage, access, and reuse of the data in research contexts.

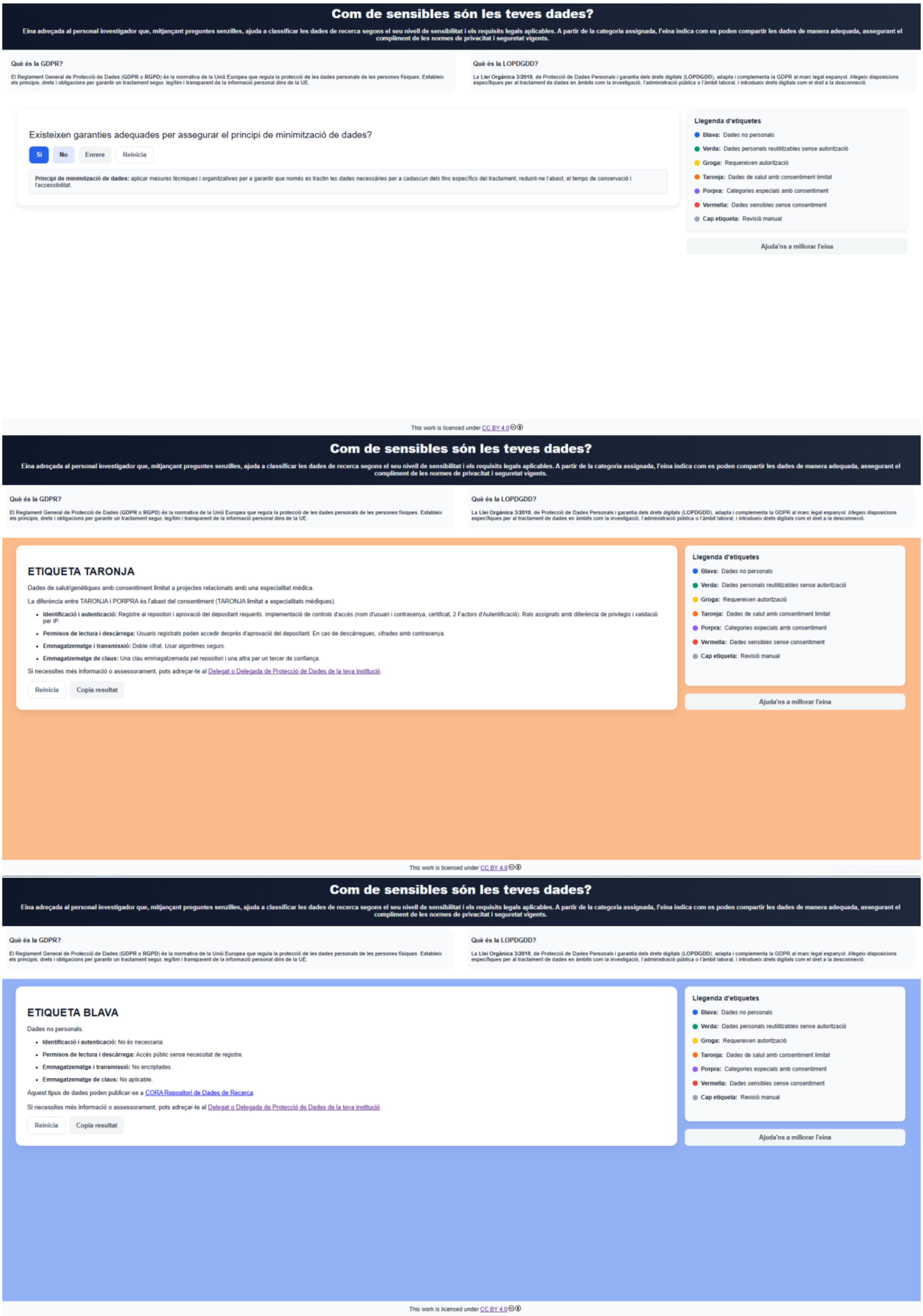

**Figure 2:** Screenshots of the interactive classification tool developed to operationalize the decision tree. The tool is currently deployed in Catalan for the local research community at the University of Barcelona.